# A Hybrid of Adaptation and Dynamic Routing based on SDN for Improving QoE in HTTP Adaptive VBR Video Streaming

**Pham Hong Thinh†, Pham Ngoc Nam†††, Nguyen Huu Thanh†, Alan Marshall††††, Truong Thu Huong†**

†Hanoi University of Science and Technology, Hanoi, Vietnam
††Quy Nhon University, Binhdinh, Vietnam
†††Vin University, Vietnam
††††University of Liverpool, United Kingdom

**Summary**
Recently, HTTP Adaptive Streaming (HAS) has received significant attention from both industry and academia based on its ability to enhancing media streaming services over the Internet. Recent research solutions that have tried to improve HAS by adaptation at the client side only may not be completely effective without interacting with routing decisions in the upper layers. In this paper, we address the aforementioned issue by proposing a dynamic bandwidth allocation and management architecture for streaming video flows to improve users' satisfaction. We also introduce an initial cross-layer hybrid method that combines quality adaptation of variable bitrate (VBR) video streaming over the HTTP protocol at the client side and SDN-based dynamical routing. This scheme is enabled by the Software Defined Networking (SDN) architecture that is now being considered as an emerging paradigm that disassociates the forwarding process from the routing process. SDN brings flexibility and the ability to flexibly change routing solutions, in turn resulting in dynamically improving the services provided in the application layer. Our experimental results show that the proposed solution offers significantly higher overall bitrates as well as smoother viewing experience than existing methods.

## 1. Introduction

In recent decades, video streaming over HTTP has become a main source of Internet traffic. A report from Cisco [1] forecasts video traffic will account for roughly 82% of the total network traffic by 2021. To cope with such enormous video traffic demand, new delivery solutions are currently being investigated in order to satisfy the expectations of the Internet users and increase revenue for network operators and media content providers.

One of the solutions considered for inclusion as a new ISO/IEC MPEG and 3GPP standard for HTTP streaming, is Dynamic Adaptive Streaming over HTTP (DASH) [2]. DASH is designed to improve the viewing experience by delivering videos with multiple and adaptive bitrates and resolutions that best match the bandwidth.

HTTP adaptive streaming is based on a request-response mechanism between clients and servers. At the server side, the media content is stored under different adaptation sets which contain video sets, audio sets, or text sets, and each adaptation set includes representations of different bitrates and types of services. A representation comprises one or more segments that contain media content and/or metadata to decode and present the streaming content. The choices of segment duration may vary from seconds to minutes following different criterion. Video segments are normally set to fixed durations from 2 to 10 seconds. After each segment has been downloaded, the client selects a suitable content version for the next segments and then makes requests to the server [3]. The advantages of long segment duration are fewer requests and less overhead, leading to higher overall throughput. However, the client can only adapt to a network change whenever it receives a full video segment, which causes a slow response time and buffer instability. Long segment duration also results in longer delays [4]. Hence, using short segment durations is a straightforward choice which imposes a significant increase in the number of requests causing an increase in overheads. This also adds to the processing complexity in network nodes and reduces overall throughput.

Currently, many adaptive strategies for HTTP video streaming have been proposed. Considering the client side, adaptive methods can be categorized into throughput-based and buffer-based methods. The throughput-based methods decide the bitrate based only on the estimated throughput and they are distinguished by different estimations and use of throughput information [5]. The buffer-based methods decide the next segment's bitrate mainly based on the buffer characteristics and the different response actions.

As a typical throughput-based method, in [6], an algorithm based on the available network throughput was proposed to dynamically adapt video quality. The algorithm used a prototype of HTTP streaming client that follows MPEG-DASH standards to evaluate the performance. In [7], a throughput-based method that is more stable to small fluctuations and responsive to large fluctuations was built to adapt to sudden bandwidth drops. However, the abrupt changes in video bitrate may cause negative effects to the useres' QoE. The problem can be solved by a buffer-based method in [8] that can provide smoother transition at sudden bandwidth drops. This method develops a trellis that represents the possible changes of bitrate and corresponding buffer level in the near future. Based on the trellis, the authors present a heuristic method to decide the bitrates of future segments while still guaranteeing that the buffer is never emptied or overrun.

The selection of representation by taking segment bitrates into account was presented in [9]. In which, the authors proposed a segment-aware rate adaptation (SARA) algorithm that considers the segment size variation in addition to the estimated path bandwidth and the current buffer occupancy to accurately predict the time required to download the next segment. In a previous study [10], we proposed an adaptive approach for video streaming over HTTP, but it only focused on CBR (constant bitrate) videos where segment bitrates throughout video are constant. Aside from the variation of network characteristics, considering VBR (variable bitrate) video cases in this paper, we also need to take into account the fluctuation of segment bitrates during playback time. The methods proposed in [5], [11], [12] achieve bitrate adaption for VBR videos that provides best effort to acquire high bitrate and better QoE in all contexts.

From another perspective, the new networking paradigm - Software Defined Networking (SDN) [13] has quickly developed and is gradually being implemented as a prominent structure for future of network technology. The most prominent characteristic of SDN is that the control plane and data plane are decoupled which provides more programmability to network implementation and development. Functionalities in SDN are performed via OpenFlow (OF) protocol [14]. Developed as an open standard protocol, OF implements flow-based switching in the OF switches to make them forward packets based on the flow-tables, which are calculated and provided by a centralized OF controller. Since the OF controller has the global view of the network, it can adjust bandwidth provisioning schemes adaptively and utilize the network resources more efficiently. Moreover, in a general case of traffic processing, SDN is robust enough not to cause network slowdown.

In [15], an SDN based dynamic traffic shaping technique for HTTP-based video streaming has been proposed. It employed an optimization model aiming to obtain maximum throughput for DASH services by selecting the optimal paths for video packet flows over SDN but didn't given the adaptation of the video bitrate. Dutra et al. [16] proposed a solution that enables the end-to-end Quality of Service (QoS) based on the queue support in OpenFlow, allowing an operator with a SDN-enabled network to efficiently allocate the network resources according to the users' demands. However, the authors addressed only the perspective of routing in which multi-paths rouing based on SDN is used. In [17] proposed a genetic algorithm-based routing method for enhanced video delivery over SDN, named GA-SDN. This work tried to improve video delivery from the rouing aspect.

There are some studies proposed in the literature for HAS over SDN which propose hybrid approach with bitrate adaptation and dynamic rerouting. In [18], the authors proposed a SDN architecture to monitor network conditions of streaming flow in real time and dynamically change routing paths using multi-protocol label switching traffic engineering to provide reliable video watching experience. In SDNDASH [19], Bentaleb et al. relies on an SDN-based management and resource allocation architecture with the goal to maximize the QoE per user considering heterogeneous QoE requirements. Each user's adaptation logic is then based on a combination of optimal bit rate recommendations and buffer levels. As an extension to this work, the authors of [20] proposes a more scalable architecture, called SDNHAS, which estimates optimal QoE policies for groups of users and requests a bandwidth constraint slice allocation, while providing encoding recommendations to HAS players. Moreover, [21] considers caching, and proposes an SDN-based Adaptive Bit Rate architecture, where video users are informed regarding each cache's content as well as get a short-term prediction of the bottleneck bandwidth to reach each cache, so that their adaptation decisions are better. In [22], Liotou et al. proposes a programmable QoE-SDN APP, based on the openness and flexibility provided by the SDN paradigm. This QoE-SDN APP can serve the customers of VSPs, improving their QoE by reducing the occurrence of the highly undesirable stalling events. Focusing on HAS applications, and by running a mobility forecasting and rate estimation function within the Mobile Network Operators domain, the proposed scheme manages to significantly improve the QoE of video streaming users. However, these studies only present the general adaptation mechanism.

Comparison of state-of-the-art studies with the proposed method is shown in Table 1.

From the literature, it would appear that solutions proposed for dynamic video streaming over HTTP have mostly solved adaptation issues at the client side only. Hence, a new adaptation requirement from the client side needs to be considered at the network side for VBR services. Therefore, in this paper, we use a SDN based dynamic routing to combine with the bitrate adaptation over HTTP as described in the proposed architecture to improve streaming VBR video quality, which subsequently leads to improved user's experience. In this architecture, an adapted bandwidth required by a client at the application layer is input to a routing decision at the network layer. Therefore, our proposed architecture can be considered as an initial cross-layer interaction model for adaptive video streaming over HTTP that synchronizes bit rate adaptation requirements at the client side and routing decisions in the transport network. Our architecture and solution also consider criteria/metrics of the video quality that influence users' satisfaction with the provided video streaming service.

Table 1: Comparison of state-of-the-art studies with the proposed solution

| Solution | Approach | Bitrate Adaptation | Adaptation Routing | Kind of Video |
|---|---|---|---|---|
| [3], [5], [7], [9] [11], [12] | HTTP | Yes | Conventiona l | CBR/VBR |
| [19]–[22] | Hybrid | General | General | CBR/VBR |
| Propose d | Hybrid | Specific | Flexible rerouting | VBR |

The rest of the paper is organized as follows: Section 2 focus describes the proposed VBR adaptation algorithm cross-layered with SDN-based routing. The performance evaluation is presented in Section 3. Meanwhile, the experiment setup, simulation results and results analysis are also described in this section. Finally, conclusion and possible future extensions are presented at the end of the paper.

## 2. Problem Formulation

### 2.1 Quality of Experience Influence Metrics of Video Streaming

Quality of Experience (QoE) is the degree of delight or annoyance of the user of an application or service. It results from the fulfillment of his or her expectations with respect to the utility and/or enjoyment of the application or service in the light of the user's personality and current state [23].

Based on the proposed architecture, the routing solution developed in the centralized controller manages network resources more intelligently in order to improve the user's satisfaction. A client selects an optimal bitrate and a corresponding 'version quality' that can enhance the viewer's QoE while distributing the available bandwidth between active streaming flows based on the VBR adaptation algorithm. According to [24], we can offer some performance evaluation metrics of video streaming on end-user satisfaction including:

- *Average Quality Bitrate*: represents the total average quality bitrate of the downloaded video segments. One of the objectives of our VBR adaptation algorithm is to maximize the average bitrate of the streamed video. For a comprehensive QoE representation, we need to combine this metric with the *Number of Version Switch-downs* explained below.
- *Number of Version Switch-downs*: represents the total number of times that a following downloaded segment has a lower bitrate than the previous segment. This metric is used together with *Average Quality Bitrate* to offer quantitative inferences about the perceived quality. If video streaming flows have the same *Average Quality Bitrate*, the flow with the lower *Number of Version Switch-downs* will be perceived better by the viewer.
- *Largest switch-down step*: The biggest downgrade in terms of two consecutive segments' bitrates over the entire streaming session. If the step is large enough, there would be a perceptible abrupt change in video quality.
- *Video Buffer*: A video (or screen/regeneration) buffer is a portion of a physical memory at the client that is used to store temporary video data. If the buffer is empty, the playback of the video has to be interrupted until enough data for playback continuation has been received. These interruptions are referred to as stalling or rebuffering. Stalling is the dominating factor of the QoE for online video streaming.

Based on the relationship between the condition of the network and the video properties, we propose an adaptation algorithm for streaming VBR video over HTTP and SDN which is used in the decision engine to flexibly request for the suitable bitrate for a segment. Our proposed algorithm aims to achieve the following targets:

1) Avoiding playback interruptions.
2) Minimizing the number of version switches.
3) Selecting optimal network path.
4) Providing acceptable quality level to make better use of bandwidth utilization.

### 2.2 Variations of Throughput and VBR Bitrate

The main distinguishing factor between the different adaptive methods lies in their adaptation logic. In the case of CBR video, each representation has almost the same bitrate for the entire period. However the bitrate of VBR video often fluctuates widely during some scene changes. Aside from the variation of network characteristics, considering VBR video cases, we also need to take into account the fluctuation of segment bitrates during playback time. Fig. 1 shows an example of the bitrates of different video versions encoded in VBR mode. This characteristic makes delivering the best video quality to viewers a real challenge. Table 2 indicates some symbols and their descriptions that are used in our study.

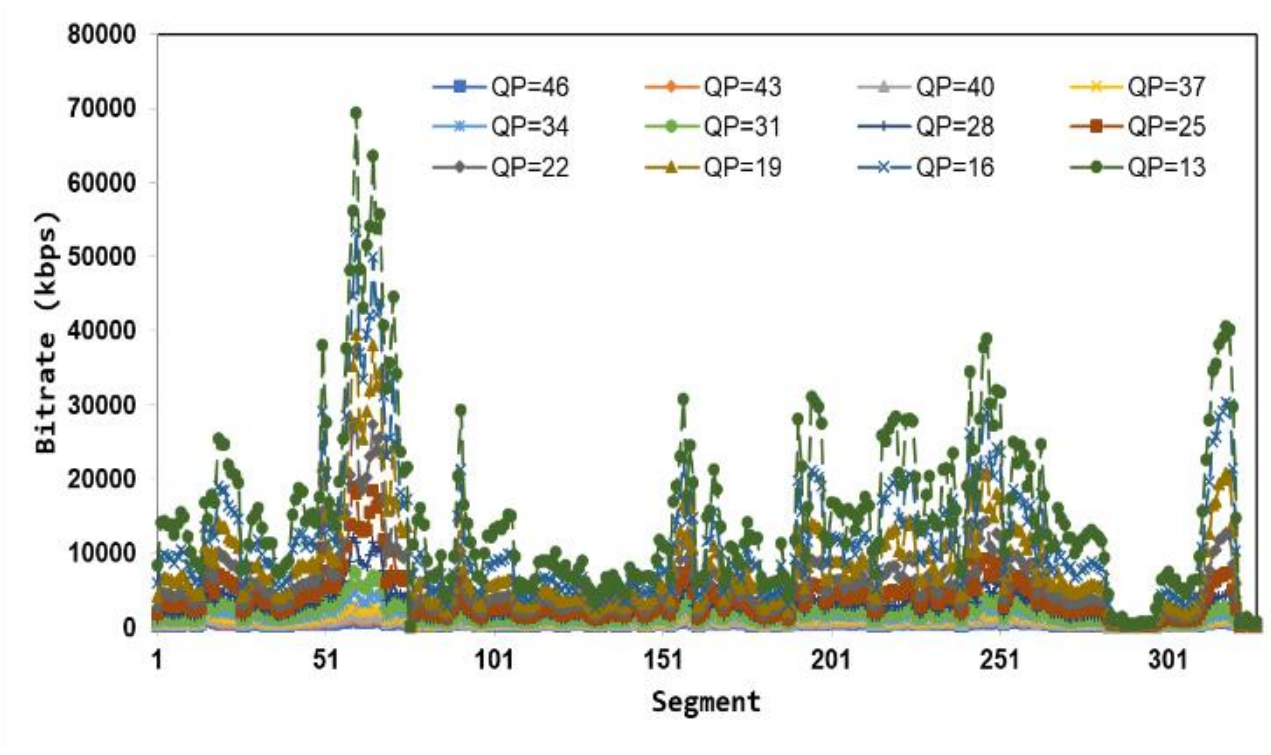

Fig. 1 Bitrates of the video versions.

Table 2: Symbols using in the paper

| Symbol | Description |
|---|---|
| $T_i$ | The throughput measured for downloading segment *i* |
| $B_i$ | The current buffer level |
| $B_{low}$ | The low buffer threshold |
| $B_{high}$ | The high buffer threshold |
| $B_{max}$ | The maximum buffer size |
| $B_{th}$ | The adaptable buffer threshold |
| $R_i$ | The bitrate of the segment *i* |
| $R_i^{avg}$ | The average bitrate of the representation of segment *i* |
| $I_i$ | The index of the representation of segment *i* |
| $R_i^{opt}$ | The optimal bitrate of the representation of segment *i* |
| $P_{i+1}$ | The optimal path re-routing |

As in [11], [12], the authors considered the buffer factor only and used it to predict the future buffer size. However the paper does not examine the related behavior of throughput and segment bitrate which is the main difficulty in VBR video streaming adaptation. To deal with the issue, we propose a simple way of examining the relationship between throughput $T_i$ (as a network property) and segment bitrate $R_i$ (as a video content property) to decide when it is necessary to change the video quality. The method can be characterized by a *deviation* parameter $\delta$, which is defined as in [25]:

$$\delta = \frac{T_i - R_i}{R_i} \quad (1)$$

From the above formula, it can be seen that the absolute value of $\delta$ implies the amount of difference between the network condition and the video quality. Furthermore, a positive value of $\delta$ states that the network is able to deliver higher video quality, while a negative value of $\delta$ means that if the client wants to avoid playback interruptions, it needs to downgrade the video representation.

To determine the representation for the next request, it is necessary to estimate the throughput based on the throughput history of received segments. Specifically, we adopt the method presented in [7], [26] where the average throughput of some downloaded video segments $T_i^s$ is used as the estimated throughput.

$$T_i^s = \begin{cases} (1-\gamma)T_{i-1}^s + \gamma T_{i,} & \text{if} \quad i > 1, \\ T_i, & \text{if} \end{cases} \quad (2)$$

where $\gamma$ is a weight in the range of [0, 1].

Moreover, since a VBR's video bitrate varies among segments, we need to use the average bitrate of the representation in conjunction with the estimated throughput to elect the optimal version. This optimal bitrate for segment $i+1$ can be defined as:

$$R_{i+1}^{opt} = Max\{R_i^{avg} \mid R_i^{avg} < (1-\mu)\ T_i^s\} \quad (3)$$

where $\mu$ is a safety margin in the range of [0, 1].

The equation means that the optimal bitrate is the specific representation's average bitrate, which is lower than the estimated throughput considering the impact of $\mu$.

## 2.3 Proposed VBR Adaptation algorithm cross-layered with SDN-based Routing – VASR

In this section, we will elaborate the so called VBR adaptation algorithm cross layered with SDN-based routing - VASR to improve the performance of the whole service delivery system. To perform VASR, we actually work on two sub packages: how to adapt bit rate at the client side, and how to route a new path based on the client new request at the nework side.

In general, VASR, with the use of a flexible threshold, our method divides the segment buffer into four ranges from empty

**Algorithm 1: VASR**

**Input:** $R_i^{opt}$, $T_i$, $R_i$, $B_i$, $\delta$, $R_i^{avg}$
**Output:** $I_{i+1}$, $P_{i+1}$

```
//Switch-up case
if B_i ≥ B_high then
    if δ > δ_0 ∧ R_i^avg < R_{i+1}^opt then
        I_{i+1} = I_i + 1;
    else
        I_{i+1} = I_i;
    end if
//Stable case
else if B_i ∈ [B_th, B_high) then
        I_{i+1} = I_i;
    end if
//Switch-down case
else if B_i ∈ [B_low, B_th) then
        if δ < −δ_0 ∧ (R_i^avg > R_{i+1}^opt ∨ R_i > R_i^avg) then
            I_{i+1} = I_i − 1;
        else
            I_{i+1} = I_i;
        end if
    end if
//Assisted switch-down case
```

20: **else if** $B_i < B_{low}$ **then**
21: **Request for a new routed path.**
22: for ($v = I_i$; $v \geq 1$; $v$--)
23: if $R_i^{avg} < R_{i+1}^{opt}$ then
24: $I_{i+1} = v$
25: end if
26: **end for;**
27: **end if**
28: **end if**

to its maximum capacity ($0 < B_{low} < B_{th} < B_{high} < B_{max}$). These ranges of buffer level are corresponding to four cases: *switch-up, stable, switch-down* and *assisted switch-down*. The details of VASR are shown in Algorithm 1.

As we can see in Algorithm 1, the whole VASR solution comprises of two strategies in combination: A VBR adaptation to adapt bit rates at the client side and an SDN-based routing.

### 2.3.1 Our proposed VBR adaptation

The first case is the *switch-up* case, which is determined by the following conditions: $B_i \geq B_{high}$. This is the case that informs the decision engine to prepare to increase the quality of the video. Nonetheless, the client only switches the requested version up if it recognize a secure deviation value ($\delta > \delta_0$), and the average bitrate of the current representation still does not exceed the favorable bitrate ($R_i^{avg} < R_{i+1}^{opt}$).

The *switch-down* case ($B_{low} \leq B_i < B_{th}$), is crucial to assure a sufficient buffer level so that it is not drained when throughput as well as segment bitrate behaves in a bad manner. So in this case, based on the information of buffer, the network must find a new route for the stream that meets the requirements. And in our proposed solution, the routing scheme is implemented in the control plane of the SDN-based system which will be elaborated in Section 2.3.2. Following our previous work [10], which presents the method to solve the same issue in the CBR case, we realized it is also effective in the VBR case. When there is a significant difference between the instant throughput and the instant segment bitrate, the dynamic buffer threshold $B_{th}$ should be closer to the high buffer level $B_{high}$. The threshold value is determined as in [25]:

$$B_{th} = B_{high} - \frac{1}{1 + e^{-\delta}}(B_{high} - B_{low}) \quad (4)$$

Similar to the *switch-up* case, only when a severely negative deviation value is detected ($\delta < -\delta_0$), and the current version is still higher quality than the optimal ($R_i^{avg} > R_{i+1}^{opt}$), or the current scene is intensive i.e. consumes extravagant network resource ($R_i > R_i^{avg}$), the client decreases the quality requested.

The *stable* case is triggered when the current buffer level is in the range of $[B_{th}, B_{high})$. This range is considered to be safe, so the client keeps the same segment representation for the next request. In the *switch-up* case and the *switch-down* case, if some conditions are not satisfied, the video quality is also maintained.

The last case, *assisted switch-down*, which is a special *switch-down* case, is indicated by the condition $B_i < B_{low}$. It means that the current buffer level is in the threatened zone. Unlike the normal switch-down where the client decreases the segment representation step-by-step i.e. one representation per request, the decision engine forcefully downgrades the video quality to an adequate figure, usually the best representation that the network can handle ($R_i^{avg} < R_{i+1}^{opt}$) in this case.

The word "assisted" implies that there is supportive cooperation supported by routing over the SDN technology. The streaming service informs the SDN controller about the drop of video quality. The client first sends a request for the next video segment to the server, then its status and the segment's properties are transferred to the controller. The controller then decides a network path considering the bitrate of the requested segment and the available bandwidth as discussed in Section 2.3.2 below. After detecting a path with improved delivering capability, the controller installs new flow rules so that the video packet can travel along that path. All of the rerouting procedures appear transparent to the client.

### 2.3.2 Our proposed SDN-based Routing

As we can see in our VBR adaptation algorithm, when a client request a new bandwidth, if the available resource of the current path does not accommodate this new bandwidth, then a new path is supposed to be found by the SDN transporation system (line 21 in Algorithm VASR). In the following paragraphs, we will describe how routing to find a new path can be implemented in the SDN-based transportation network.

The proposed controller architecture, depicted in Fig. 2, offers a Controller-Forwarder interface and a variety of functions. The interface uses the OpenFlow protocol, providing a secure way for exchanging information between the controller and forwarders such as routers or switches. These messages include network topology discovery, flow-table modifications

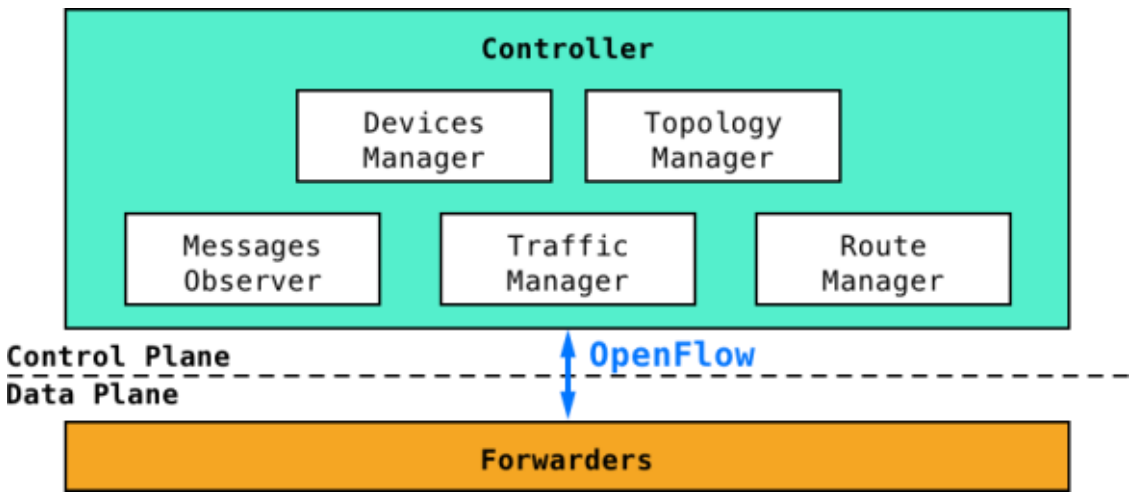

Fig. 2 Proposed OpenFlow controller and interface.

and resource monitoring. The controller's main functions are delivered by several modules:

- *Topology Manager*: This module is responsible for discovering the network topology i.e. routers, switches and links by sending check-up messages on a regular basis.
- *Devices Manager*: This module supervises hosts in the network. Each host is considered as an attachment to the forwarder with which it has direct connection.
- *Traffic Manager*: This module collects statistical data from forwarders to determine the packet forwarding performance, which is used to help the route calculation.
- *Route Manager*: This module cooperates with the Topology Manager and Traffic Manager, and performs routing algorithms to obtain specific network routes between hosts. It is also responsible for installing flow rules onto routers or switches.
- *Messages Observer*: This module handles messages from application services.

When the controller starts up, the *Topology Manager* and the *Devices Manager* start listening for connections from the OpenFlow forwarders. After exchanging messages, the network topology i.e. networks devices' configurations and links' statuses are acquired by the controller.
When a streaming session is established, the video service notifies the controller by sending an initial message, which is processed by the *Messages Observer*. At that time, the *Route Manager* determines the best-effort path for transmitting packets between the server and the client, then updates the flow table rules of appropriate forwarders. The *Traffic Manager* is also activated to monitor multimedia flows in the network. It may trigger the routing module again to find a new route according as the traffic policy.
In order to investigate the performance of our VBR adaptation algorithm assisted by a SDN-based routing scheme, we propose to check up the three different possible schemes and find the most appropriate one: SDN-based Periodical routing (SPR), SDN-based Adaptive Routing without Monitoring (SAR), and SDN-based Adaptive Routing with Monitoring (SARM).

### 2.3.2.1 SDN-based Periodical Routing (SPR)

Mechanism of implementing periodical path routing based on SDN is shown in the flowchart of Fig. 3. The controller cannot determine the condition of a link if no packets traverse that link. Our controller is implemented with a Round Robin method [27]; therefore, video packets are used to measure the approximate bandwidth i.e. maximum amount of data that can be transmitted on each path at a certain moment.
In the switching stage, each path is alternately selected every $t$ second(s). The controller stores the measured bandwidth of the previous path before sending new flow entries of the next path to the switches. When the bandwidths of all links have been estimated, the controller chooses the path with the highest bandwidth at that time to serve. Thus, the optimal bandwidth of all the paths is determined as follows:

$$BW^{opt} = \max_{p\in[1,n]}\{BW_p\} \quad (5)$$

That path is maintained for the next $\alpha \times t$ seconds before the entire procedure repeats again. The total time $T$ for a procedure is calculated by the formula:

$$T = (n+\alpha)t, \quad (6)$$

where $n$ is the total number of paths, $\alpha$ is the time to remain in a particualr path and $t$ is the switching period.
However this Round Robin method is inadequate when the number of paths between two hosts is large since the instantaneous path throughputs will not be reflected correctly after the switching stage. Also, the switching period needs to be sufficiently long enough for the controller to process the statistic queries. These queries are merely default OpenFlow messages used to exchange physical ports' information between switches and the controller. Typically, $t$ is larger than one second. Therefore, the simplest solution with round robin flow scheduling also has the most overhead in terms of control channel traffic.

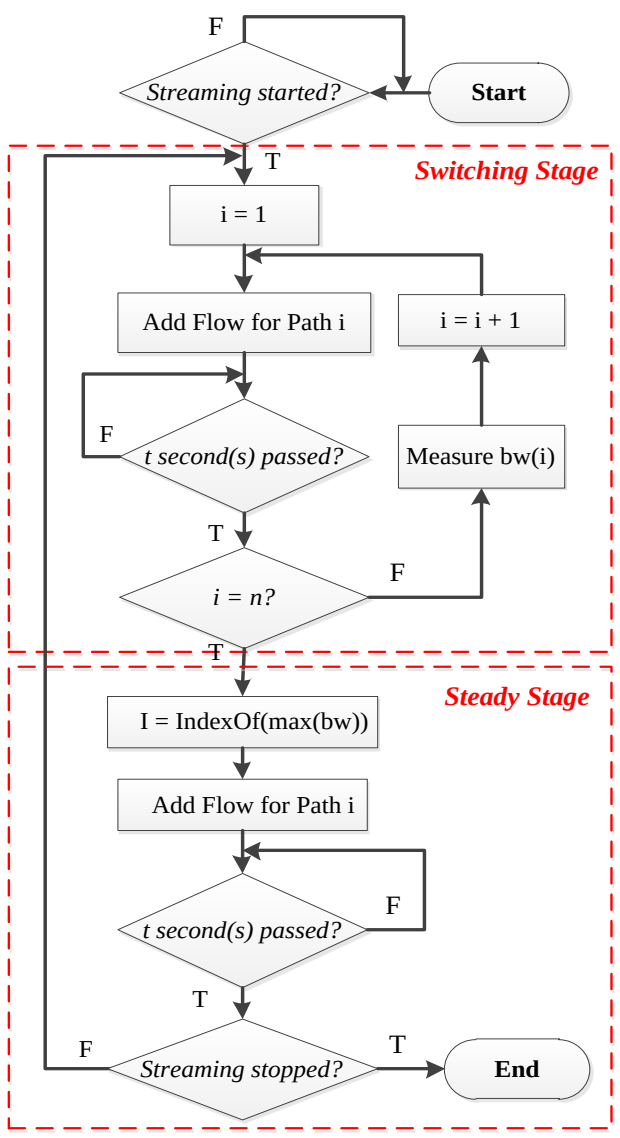

Fig. 3 The flowchart for routing.

The controller needs to periodically query the flow table counters in the switches to detect the current bandwidth utilization of each link. Additionally, due to the nature of traffic distribution, more flow redirection operations are required, which increases the number of transmitted *Flow-Mod* messages, which are defined in the OpenFlow protocol in order to allow the controller to modify the state of an Openflow switch, as well as the workload of the controller and switches.

### 2.3.2.2 SDN-based Adaptive Routing without Monitoring (SAR)

Unlike the periodical routing case, controllers which implement the adaptive routing mechanism do not actively alter the path between hosts, and work in a passive manner. The rerouting procedure is performed only when the controller is demanded by the client. In the *assisted switch-down* case, when the current buffer level is less than the low buffer threshold ($B_i < B_{low}$), the system must find out a new route/path that meets the requirement on bandwidth of the corresponding client. This rerouting policy is almost identical with the periodical one, especially the switching stage in which the controller interchanges paths in the available pool to detect the preeminent one. The difference arises in the steady stage (i.e. the length of time a particular path is chosen for). The controller keeps the preferred path until requested to reroute again in lieu of rerunning the procedure, which means less work is required from the controller. This policy is referred to in the flowchart of Fig. 4.

### 2.3.2.3 SDN-based Adaptive Routing with Monitoring (SARM)

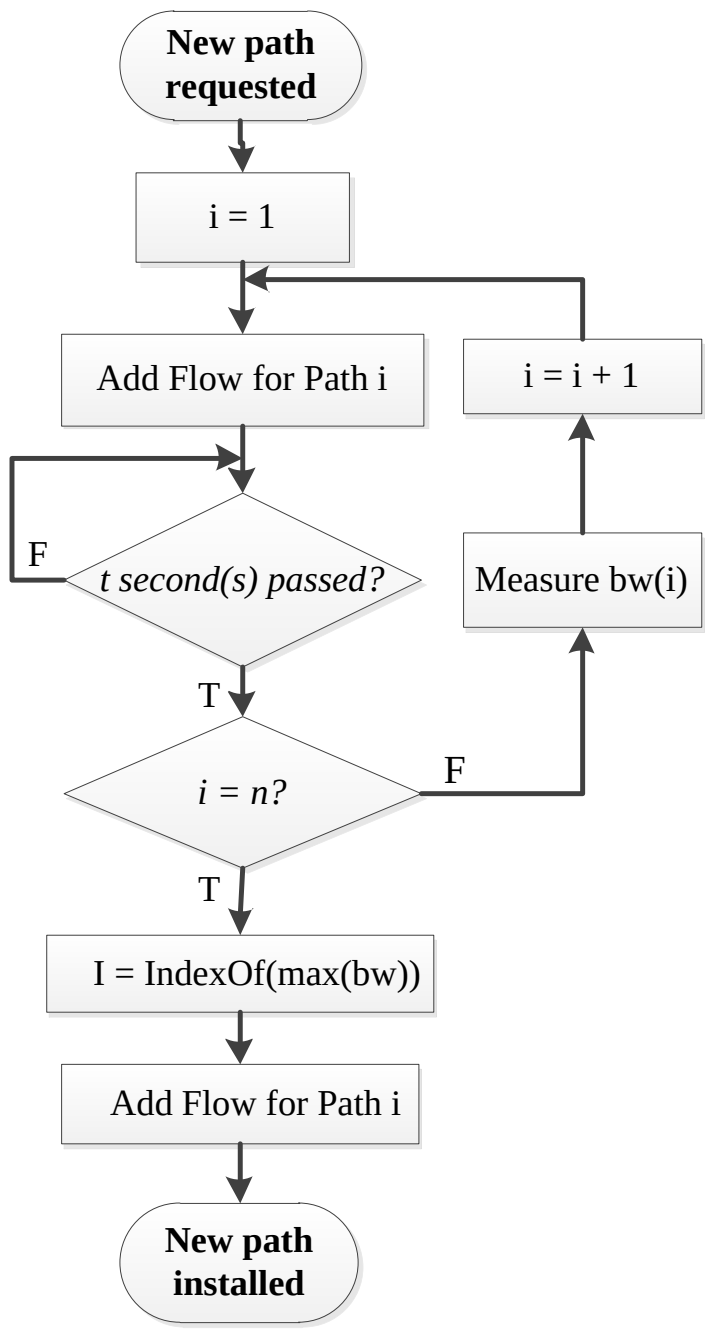

Fig. 4 The flowchart for Adaptive Routing

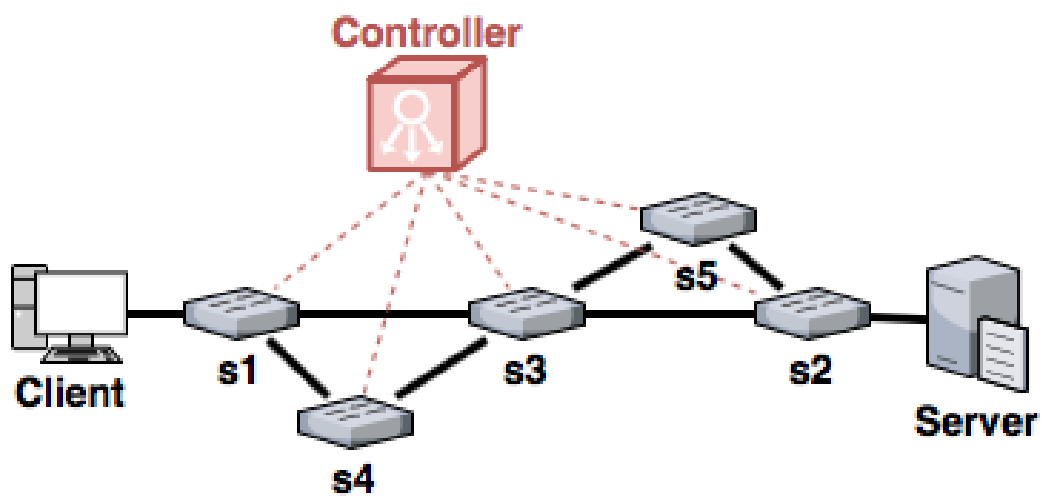

Fig. 5 Network topology of experimental testbed.

Since link congestion may occur by the time the controller is informed by the client, it is essential to run a monitoring process alongside the observing process. In the monitoring process, the controller continuously measures the throughput of the current path which is delivering the video data. Whenever the path does not meet (less than) the threshold throughput $BW_{th}$ requirement i.e. the link is deemed to be congested, and the controller intermediately seeks a new path without waiting for the client's request.

## 3. Performance Evaluation

In this section, we compare our solutions with two other algorithms: The instant throughput-based algorithm called "Aggressive" [7] and a Segment-Aware Rate Adaptation algorithm called "SARA" [9].

### 3.1 Experiment Setup

The testbed setup in this paper is illustrated as Fig. 5. The proposed algorithm is simulated in Mininet [28] to create our network topology, which has a DASH server, a DASH client and five switches ($s_j$, $j = 1 \div 5$). The server is simply an Apache Webserver version 2.4.7 running on Ubuntu 14.04. The switches create many possible paths from client to server and they were connected to a remote controller – Floodlight [29]. The controller takes the jobs of flow control, such as route calculation and selecting the optimal path.

At the DASH server side, the test video is a short animation movie named Elephants Dream [30], which is ten minutes and fifty-four seconds long. The original video is VBR-encoded using the H.264 codec into 12 versions with different quantization parameters (QPs). In the experiments, the set of QP values are 13, 16, 19, 22, 25, 28, 31, 34, 37, 40, 43, 46. All the video versions are served at Full-HD resolution (1920x1080) and 24 frames per second. Each version is divided into small video segments of 2 seconds, which means that there will be a total of 327 segments downloaded in one conducted simulation. The segment bitrates of all versions are shown in Fig. 1. The version index as well as its QP and corresponding average bitrate of the version are listed in Table 3. There is a manifest file which contains information about all the video versions [2]. The DASH client must request for this MPD (Media Presentation Description) file before the video stream starts downloading.

The DASH client is a simple video player which implements the MPEG-DASH standard by using the libdash library. Its main functionality is requesting the manifest file from the server, playing the video content and adapting video quality to the network condition. The adaptive algorithm used in this paper is a buffer-based algorithm which depends on both the throughput and the video buffer measured at the client side together dynamic routing mechanism over SDN (Algorithm 1).

As can be seen in Fig. 5, there are four possible paths for packets to be transmitted from server to client as follows: s2 – s3 – s1; s2 – s3 – s4 – s1; s2 – s5 – s3 – s1; s2 – s5 – s3 – s4 – s1.

To manipulate the fluctuating bandwidth of network links in reality, we use a Traffic Control (TC) [31] technique on downlink network interfaces along each path. Traffic Control works on packets *leaving* the system. The TC code operates between the IP layer and the hardware driver that transmits data on the network. This means that the TC module i.e. the packet scheduler is permanently activated in the kernel, even when it is not explicitly required. By default, this scheduler maintains a basic *First-In First-Out (FIFO) queue* in which the first packet arrived is the first to be transmitted. At the core, the TC is composed of *queuing disciplines*, or *qdisc*, that represent the scheduling policies applied to a queue. In this case, we implement a Token Bucket Filter (TBF) that assigns tokens to a *qdisc* to limit it flow rate.

To perform the experiment to select the optimal path, we assume that the bandwidth on the paths is different and evaluate the bandwidth traces of the four paths as in Fig. 6. The red line (Best Path) is an imaginary boundary which highlights the best possible link capacity at a certain moment. The full bandwidth simulation is actually longer, but after conducting several experiments, we found that the last video segment would be downloaded by the 560-second timestamp; therefore, we trim the traces graph for visualization purposes.

The dynamic routing mechanism based on SDN and described in Fig. 3 and Fig. 4 is used. The flowchart of Fig. 3 can divide into two states: the switching state and the steady state. The parameters in Equation 6 are fixed setup, $n = 4$, $\alpha = 1$, $t = 2$s.

Table 3: Version information of the test video

| Version Index | QP | Average Bitrate (kbps) |
|---|---|---|
| 0 | 46 | 354 |
| 1 | 43 | 472 |
| 2 | 40 | 638 |
| 3 | 37 | 882 |
| 4 | 34 | 1.234 |
| 5 | 31 | 1.779 |
| 6 | 28 | 2.588 |
| 7 | 25 | 3.823 |
| 8 | 22 | 5.613 |
| 9 | 19 | 8.028 |
| 10 | 16 | 11.156 |
| 11 | 13 | 15.227 |

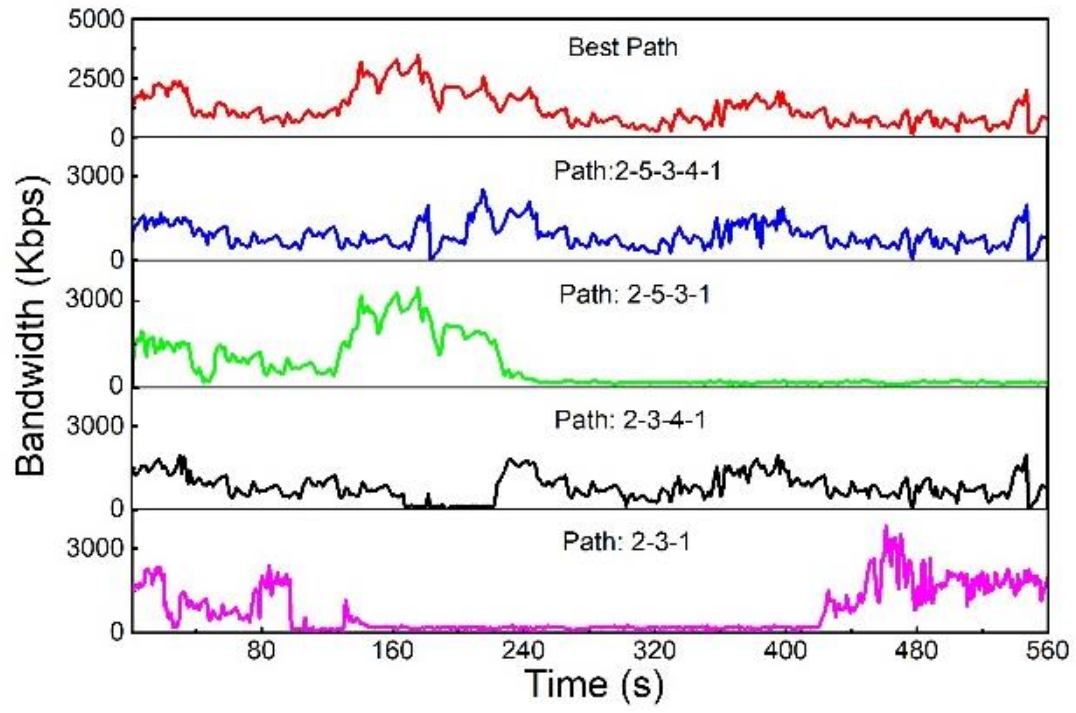

Fig. 6 Bandwidth scenarios in four paths of the network.

## 3.2 Scenarios and Experiments Description

In order to evaluate the performance of different implementations of the SDN controller's routing policy, we introduce three scenarios that cover six experiments:

- ✓ *Scenario 1*: The controller is activated with its default behavior, which turns the OpenFlow switches into traditional learning switches. In this type of network, the best-effort path between two hosts is usually the path with minimum number of hops.
- ✓ *Scenario 2*: The controller is implemented with the periodical routing policy. In this case, it ignores the routing request from the media player, and operates independently for the entire streaming session.
- ✓ *Scenario 3*: The controller is implemented with the adaptive routing policy. The monitoring process may or may not be enabled.

Among these three scenarios, we conduct a total of six experiments:

- *Experiment 1*: The "Aggressive" solution is experimented with Scenario 1
- *Experiment 2*: The "SARA" solution is experimented with Scenario 1. The parameters in this algorithm are fixed setup $I$ = 10s, $B_\alpha$ = 15s, $B_\beta$ = 25s, $B_{Max}$ = 50s. Such factors are selected based on [9] and our own experiments.
- *Experiment 3*: The "SDN-based SARA" solution is experimented with Scenario 2. In this experiment, we combine the bitrate adaptation algorithm of SARA with SDN-based Periodical Routing mechanism. The switching period is equal to the video segment duration ($t = 2s$). The best path is maintained for one switching period ($\alpha = 1$). The parameters in the SARA algorithm are fixed setup $I$ = 10s, $B_\alpha$ = 15s, $B_\beta$ = 25s, $B_{Max}$ = 50s.

- *Experiment 4* (Scenario 2): The switching period is equal to the video segment duration ($t = 2s$). The best path is maintained for one switching period ($\alpha = 1$). The bitrate adaptation is used as in the proposed algorithm VASR.
- *Experiment 5* (Scenario 3): The switching period is equal to the video segment duration ($t = 2s$). The monitoring process is not enabled. The bitrate adaptation is used as in the proposed algorithm VASR.
- *Experiment 6* (Scenario 3): The switching period is equal to the video segment duration ($t = 2s$). The monitoring process is enabled ($BW_{th} = 1{,}000kbps$). The bitrate adaptation is used as in the proposed algorithm VASR.

For all cases, the number of possible paths is four ($n = 4$). At the client side, the VBR adaptation logic uses the following parameters: $B_{low} = 15s$, $B_{high} = 25s$, $B_{max} = 50s$ and $\delta_0 = 0.5$. The client always requests for the segment with lowest quality first.

## 3.3 Experimental Results and Discussion

The results of *Scenario 1* (Experiment *1* and *2)* for streaming a VBR video over HTTP protocol in a traditional switch network are shown in Fig. 7 and Fig. 8. As stated earlier, in a non-SDN network, data packets usually travel along only the path with minimum number of switch hops; in this case, the path is $s2 \rightarrow s3 \rightarrow s1$.

As may be observed, the shape of the Download Rate series is almost identical to the bandwidth of the path $s2 \rightarrow s3 \rightarrow s1$. At the end of the streaming session's first quarter (around segment 80 – by investigating the experiment's detailed log), the three-switch path suffers a severe congestion and the throughput never exceeds 500kbps. In this condition, since there is no rerouting mechanism available, the media player can only adapt by downloading poor quality versions of the video. It is not until the last quarter that the path's traffic capacity recovers, and the best representations are requested again.

The tests conducted in *Scenario 2* (*Experiment 3 and 4)* employ the Periodical Routing mechanism on the SDN controller. In all two trials, we retain the preeminent path for only one switching period in the Steady Stage. We set the switching period equal to video segment duration. Fig. 9 shows the result when running SDN-based bitrate adaptation algorithm of SARA at the client side, while Fig. 10 shows the result for our bitrate proposed adaptation algorithm case.

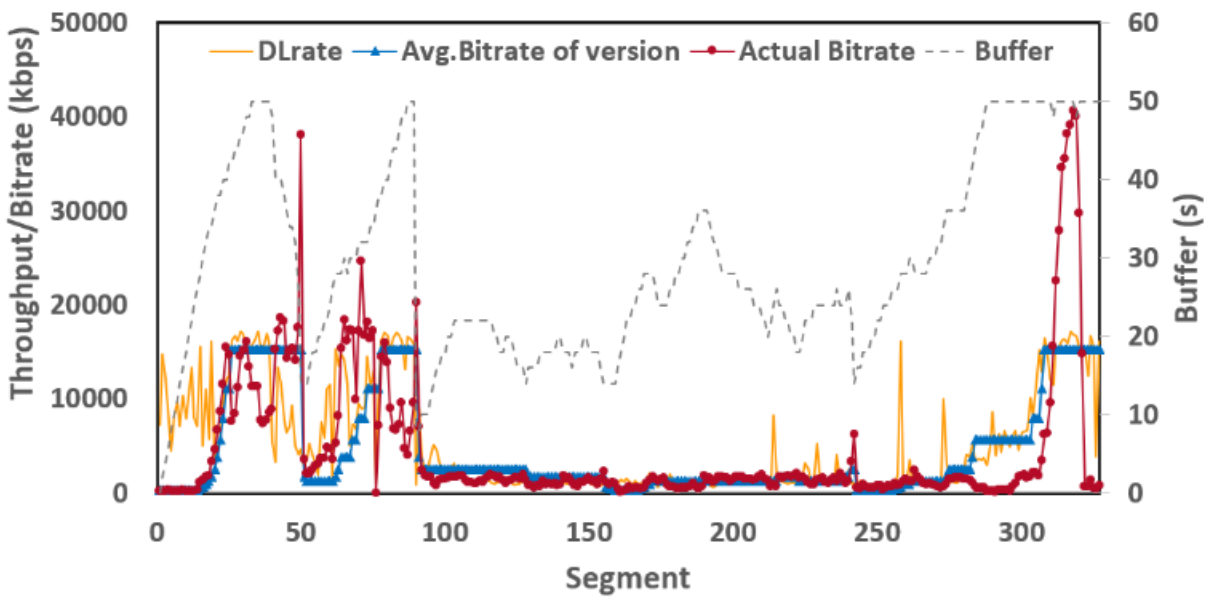

Fig. 7 Adaptation results for "Aggressive" method.

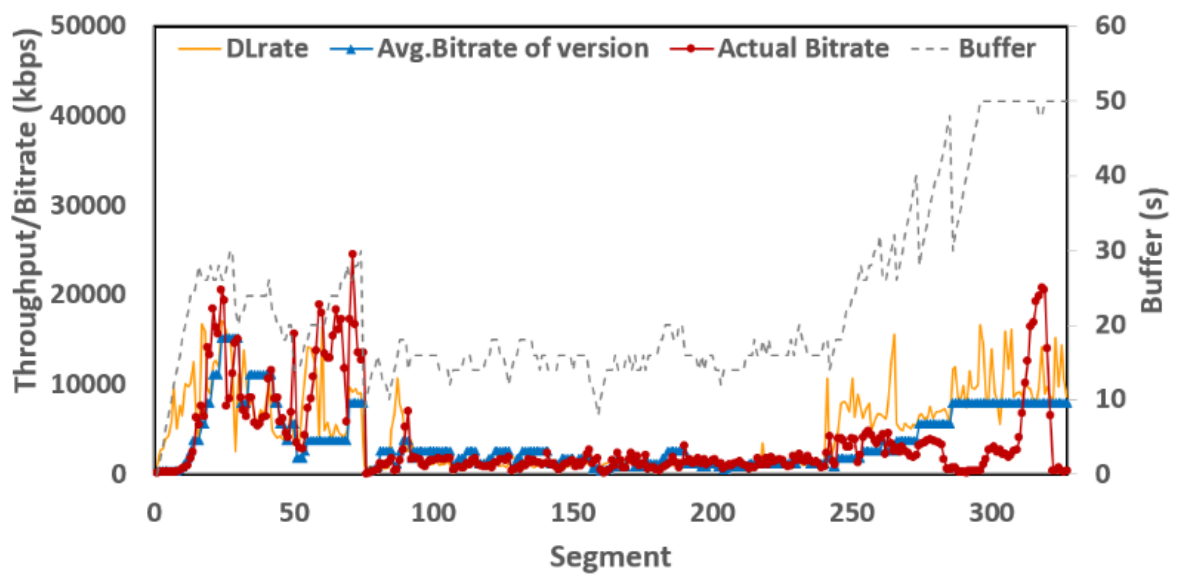

Fig. 8 Adaptation results for "SARA" method.

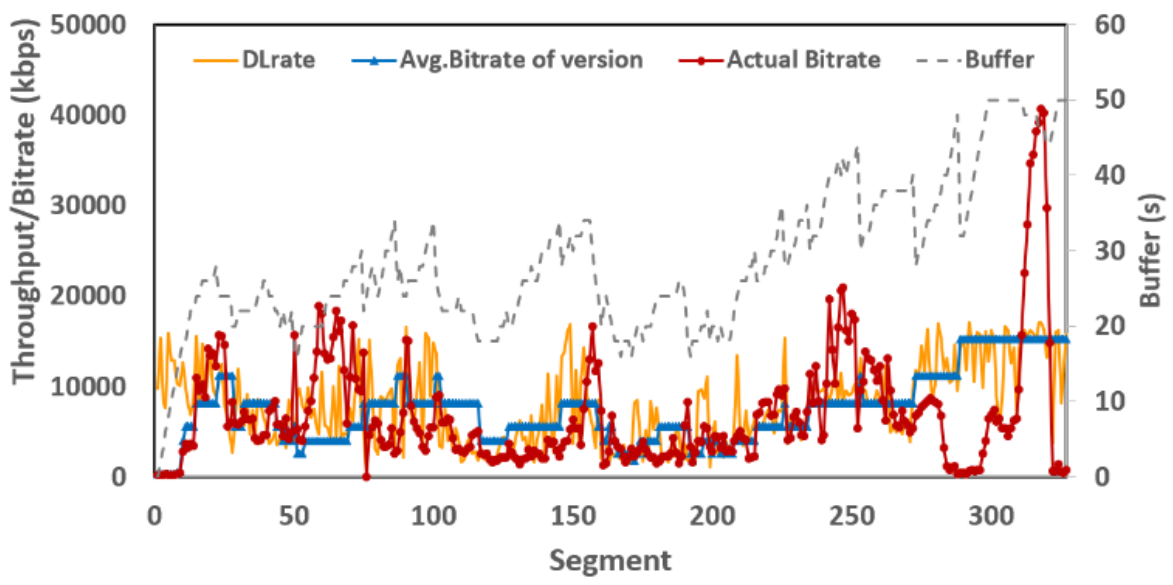

Fig. 9 Adaptation results of the SDN-based SARA.

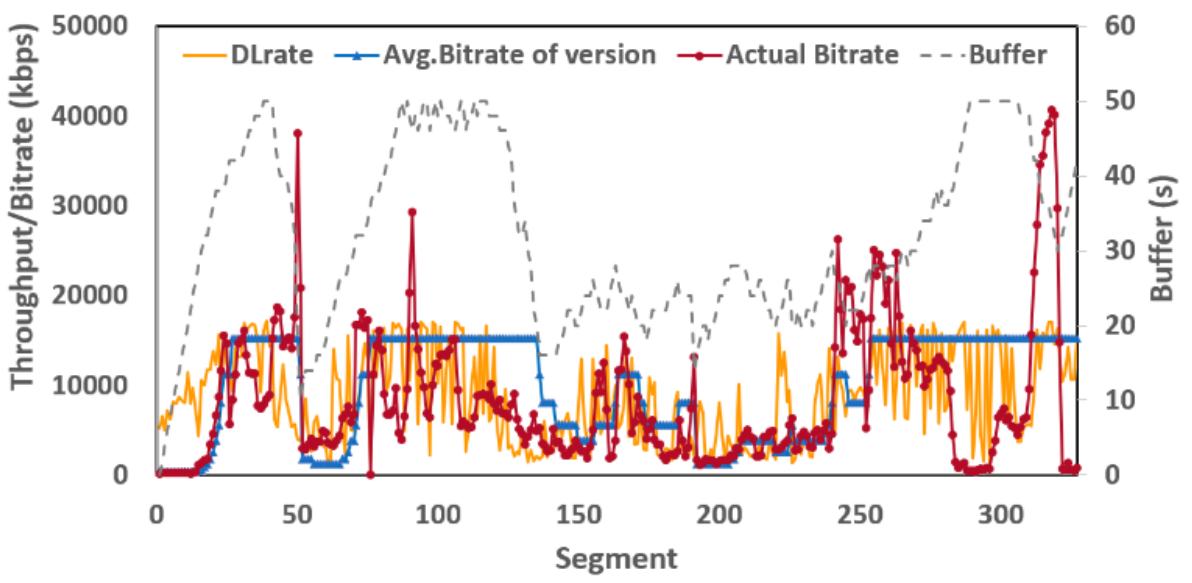

Fig. 10 Adaptation results of the periodical routing with SDN (SPR)

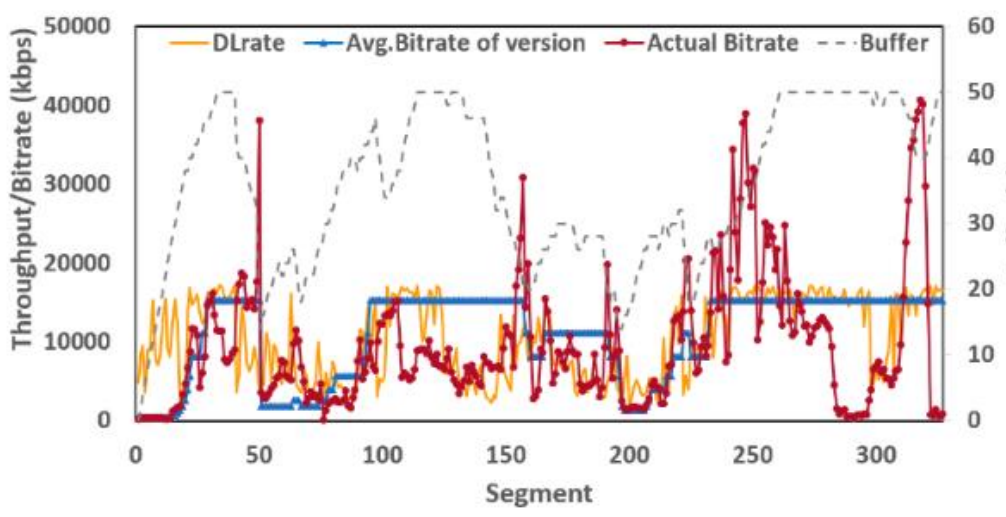

Fig. 11 Resulting of the adaptive routing without monitoring process (SAR).

The last scenario (*Scenario 3*) is the implementation of the Adaptive Routing mechanism in the SDN controller. It is divided into two cases: with and without the self-monitoring process. For the latter case (SAR), the controller computes the new path only if requested by the streaming service. Fig. 11 shows the result when the monitoring process is not present. The first and the last quarter of the streaming session look identical to the experiment without the presence of the controller, while the middle part appears to be superior. This suggests that the video data should not be downloaded on the shortest path, but via another path with better performance. For the former case (SARM), the controller computes the new path when it discovers a bottleneck in the current path. It can be seen from Fig. 12 that the best representation is downloaded most of the time. There are only two version drops at around segment 52 and segment 178, which is possibly due to path reallocation.

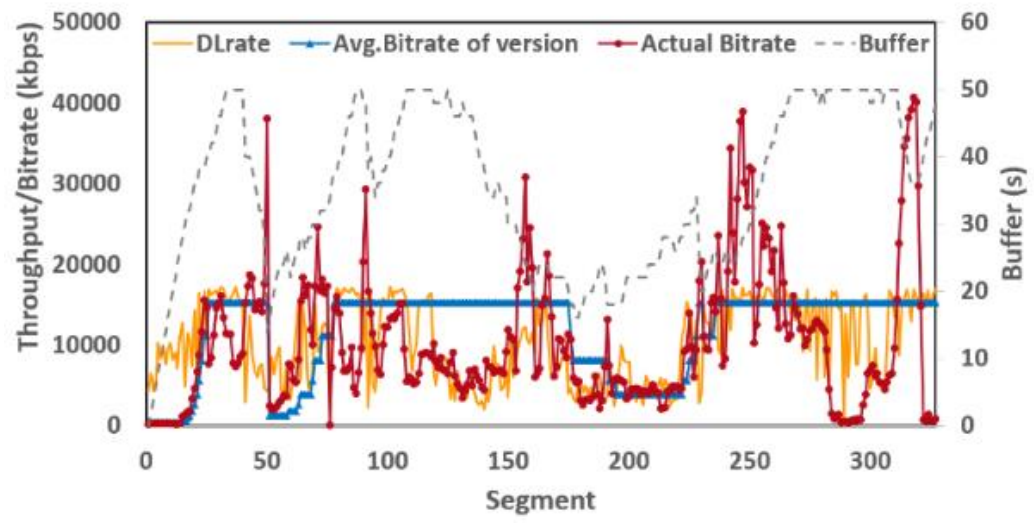

Fig. 12 Resulting of the adaptive routing with monitoring process (SARM).

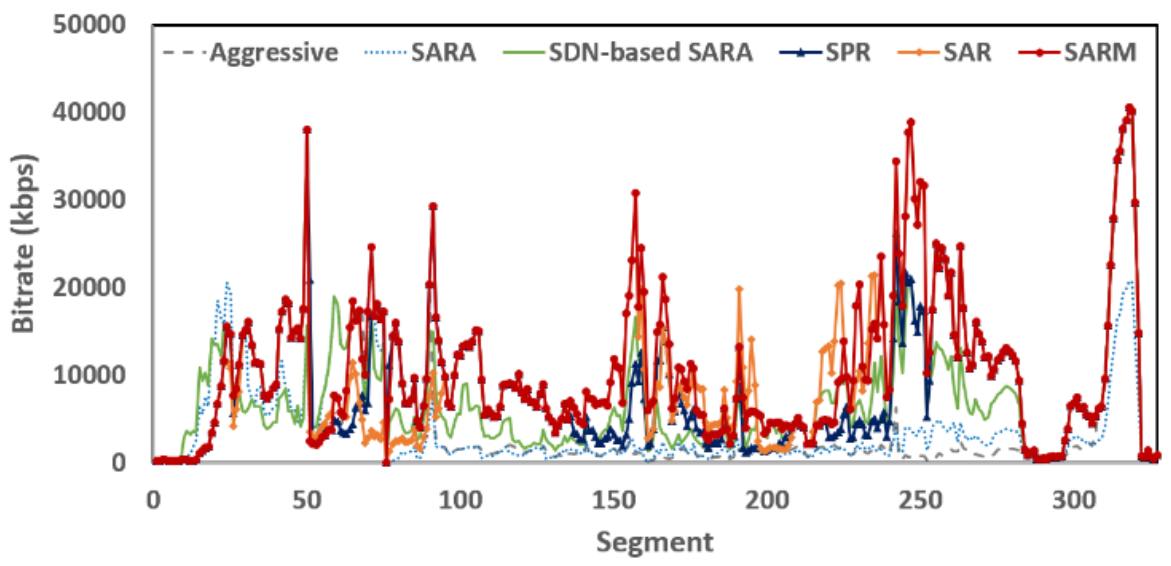

Fig. 13 Actual bitrate results of all methods.

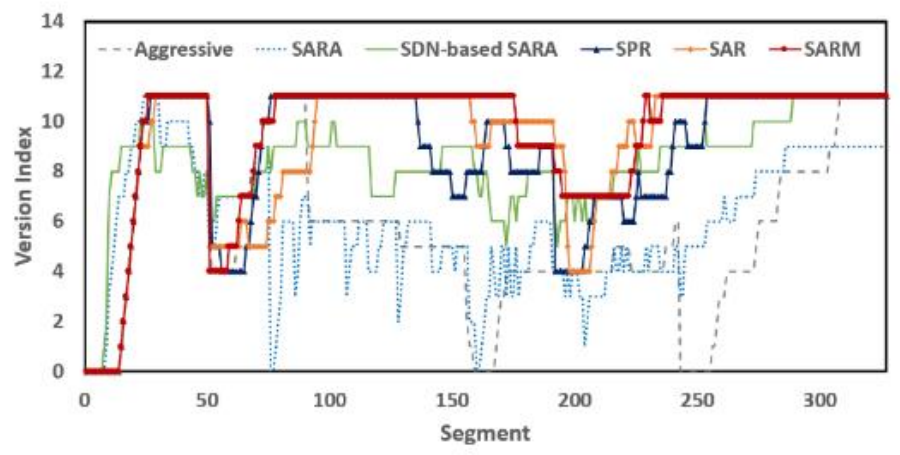

Fig. 14 Average version index results of all methods.

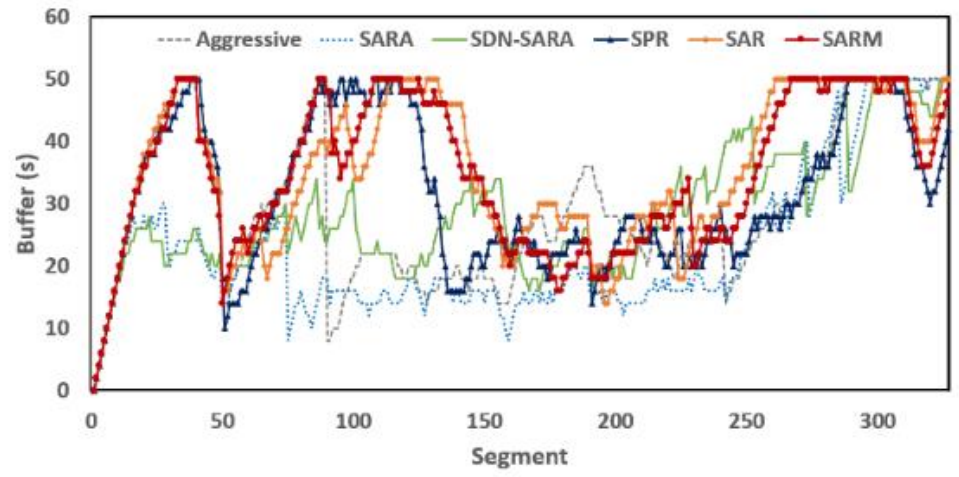

Fig. 15 Resulting buffer level of all methods.

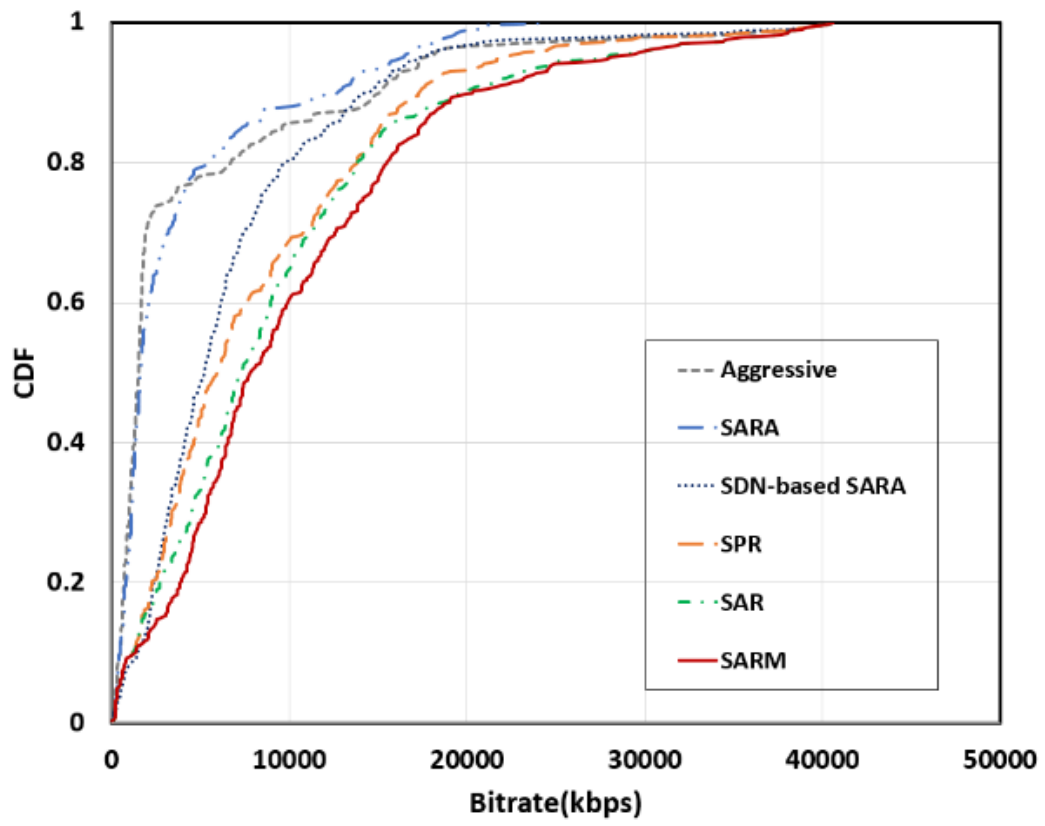

Fig. 16 The cumulative distribution function of bitrate.

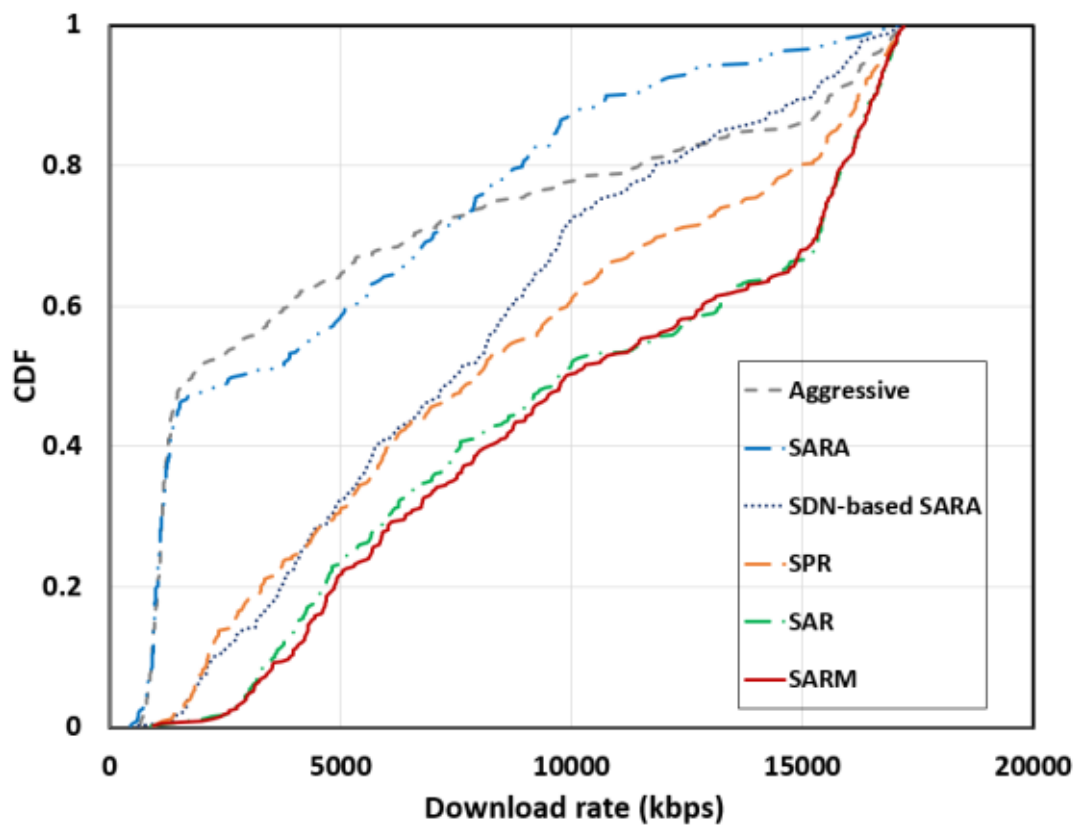

Fig. 17 The cumulative distribution function of download rate.

Table 4: Statistics of different methods

| Criteria | Aggressive | SARA | SDN-based SARA | Proposed VASR | | |
|---|---|---|---|---|---|---|
| | | | | SPR | SAR | SARM |
| Average bitrate (kbps) | 4435 | 3782 | 6864 | 8296 | 9404 | 10238 |
| Average version index | 5.7 | 5.9 | 8.3 | 8.7 | 9.1 | 9.5 |
| Average video buffer (s) | 30 | 23 | 29 | 32 | 36 | 35 |
| Proportion of buffer < 15s (%) | 7.6 | 23.2 | 2.4 | 4.5 | 3.3 | 2.7 |
| Number of version switch-downs | 10 | 32 | 22 | 13 | 9 | 6 |
| Largest switch-down step | 6 | 9 | 2 | 5 | 6 | 7 |

We plot several graphs in order to show the difference in performance among approaches. Fig. 13 depicts the actual bitrate of downloaded segments, and Fig. 14 is the corresponding average version index results. At the start of the streaming session, since our adaptation strategy at the client is buffer-based, video segments with the lowest bitrate are always downloaded until it is safe to switch to better versions. Thus, in the first 16 seconds, which corresponds to the first 8 segments, all methods are identical. When switching up versions, our decision engine performs one-by-one steps providing smooth transitions among qualities. By the time segment 25 is downloaded, all methods have reached the highest quality version. From that segment onwards, each method produces a completely different behavior. It is not until segment 270 that all methods show the same results again, the reason being that all paths in the network are sufficient for delivering highest bitrate segments at the end of the streaming session. It can be inferred from these two plots that the non-SDN experiments (Aggressive, SARA) undoubtedly produce the worst result, which could be clearly seen from the period between segment 80 and segment 270, while Periodical Routing scenarios (SDN-based SARA, SPR) show some moderate enhancements. Above all, Adaptive Routing scenarios deliver better video quality, especially with the monitoring process (SARM).

When considering the buffer level (Fig. 15), the Adaptive Routing experiments (SAR, SARM) conspicuously emerge as the leading solutions. The period from segment 20 to segment 40 and from segment 80 to segment 140, by always tracking the current links' condition, these methods are able to accumulate the video buffer while maintaining the best video quality. Other scenarios show indistinguishable buffer behavior; nonetheless, the video segments stored in each buffer are of disparate qualities. In the next period from segment 140 to segment 170, the SARM suffers a significant drop in terms of buffer level. It could be explained that during this period, the network bandwidth availability (Fig. 6) is not enough to deliver high bitrate segments efficiently i.e. it takes more time to download a segment than to play it. Overall, because the bitrate adaptation algorithm is throughput-buffer based, the buffer level in all cases exhibits large fluctuations.

The comparison using cumulative distribution functions (CDFs) of bitrate (Fig. 16) and download rate (Fig. 17) gives an insight into a client's overall behavior. The first thing to notice from the bitrate CDF figure is that the rate of segments downloaded at the high bitrate of SARM is greater than the rest. Specifically, around 50% of segments have bitrate greater than 8000kbps. The CDF value for SDN-based SARA, SPR and AR is lower about 28-46% while the figure for Aggressive and SARA is lowest at 14-17%. We can see from Fig. 17 that SDN-based proposed solutions offer a higher download rate compared to fixed path algorithm. The proportion of segments downloaded with download rate less than 5000kbps of SARM, SAR, SPR and SDN-based SARA are around 22%, 24%, 31% and 32% respectively. While the figure for Aggressive and SARA with fixed path is much higher at around 58-60%. The adaptive routing policy of proposed method enable about 50% of segments are downloaded with download rate greater than 10000kbps. However, the CDF for the other algorithms without SDN is only from 13-22%.

For a clearer interpretation, the detailed results of all methods are provided in Table 4. Since our controller operates on a computer running Linux, we sample the CPU states every one second using the Linux built-in processes manager; then, we take the average of all samples. This criterion may show different results on different setups, but the relative relation between methods should not be changed. As shown in the Table 4, the proposed methods are capable of enhancing some QoE parameters for video spectators. To investigate the result in details, we divide the scenarios into three categories: Non-SDN (Aggessive, SARA), Periodical Routing with SDN (SDN-based SARA, SPR) and Adaptive Routing with SDN (SAR, SARM). Video streamed via a software-defined network undoubtedly has superior average bitrate i.e. quality compared to a traditional network.

For the Periodical Routing cases, the overall video quality is better than the non-SDN cases. However, these methods does not guarantee the improvement of buffer levels. These methods also demands the most amount of work from the controller. In contrast, the Adaptive Routing with Monitoring (SARM) delivers the highest average bitrate of 10,238kbps, greater than twice that of the non-SDN cases (3,782kbps and 4435kbps). Additionally, the average version index, the average buffer and the fraction of low buffer status (i.e. the current buffer is less than the low buffer threshold) over the entire streaming session show the best performance among all methods. The proportion of streaming time that the buffer dropped below 15s (danger zone) in SARM is significantly low (2.7%) as compared to other methods. The only better-performed method in this criteria is SDN-based SARA, which is 2.4%. This method also records the lowest number of version switch-downs of 6, while the non-SDN cases has to switch down the video quality a total of 10 and 32 times over the entire session, which means that the video is delivered more smoothly and stably.

Another important criteria is Largest switch-down step where the bigger the step between two consecutive segments, the more abrupt is the video quality change. Such sudden changes in quality may bring discomfort and annoyance to viewers. In the case without the presence of a SDN controller, the largest step recorded is 9 for SARA, from version 9 at segment 75 down to version 0 at segment 76. However, when coupled with a controller, SDN-based SARA only produces the largest switch-down step of 2-level. In our proposed solutions which SPR, SAR and SARM, the biggest step-sizes are 5, 6 and 7 respectively, which is fairly acceptable.

In other words, the Adaptive Routing with Monitoring (SARM) appears to have the best performance one among all methods; however, if the controller utilization is taken into account, the Adaptive Routing without Monitoring (SAR) still provides acceptable enhancements in the video perceiving experience.

## 4. Conclusion and future work

With the tremendous increase in media content consumption over the last decade, especially high-definition videos over the Internet, it is essential to have a streaming architecture that can cope with highly varying delivery conditions in order to improve the users' QoE.

In this paper, we have presented a novel method for adaptive streaming of VBR videos over the HTTP protocol based on the buffer level and the estimated throughput combined with dynamic network path allocation in the context of software-defined networking. A variety of experiments have been conducted to investigate the performance of the VBR adaptive algorithm with and without the aid of SDN. The experimental results have shown that the proposed methods, especially with the SARM outperforms other existing non-SDN methods with an improvement of up to 200% in terms of delivered video bitrate.

To develop this work in the future, we intend to broaden the topology and increase the number of clients. We also plan to add heterogeneous types of clients such as HDTV, laptop and mobile users; and improve the optimization model by considering the client properties. The SDN architecture may also be extended to consider multiple shared networks and different patterns of dynamic traffic, and to integrate new bitrate decision logics, bandwidth and QoE estimators.

### Acknowledgment

This work was supported by the grant of Hanoi University of Science and Technology, code: T2018-PC-065

## References

[1] "Cisco Visual Networking Index: Forecast and Methodology, 2016-2021," San Jose, CA, USA, Cisco, White Paper, 2016.

[2] T. Stockhammer, "Dynamic Adaptive Streaming over HTTP – Standards and Design Principles," Proc. Second Annu. ACM SIGMM Conf. Multimed., no. i, pp. 133–143, 2011.

[3] T. C. Truong, H. T. Le, H. X. Nguyen, A. T. Pham, J. W. Kang, and Y. M. Ro, "Adaptive Video Streaming over HTTP with Dynamic Resource Estimation," J. Commun. Networks, vol. 15, no. 6, pp. 635–644, 2013.

[4] S. Wei and V. Swaminathan, "Low Latency Live Video Streaming over HTTP 2.0," Proc. ACM NOSSDAV, pp. 37–42, 2014.

[5] T. C. Truong, H. T. Le, A. T. Pham, and Y. M. Ro, "An evaluation of bitrate adaptation methods for HTTP live streaming," IEEE J. Sel. Areas Commun., vol. 32, no. 4, pp. 693–705, 2014.

[6] K. Miller, E. Quacchio, G. Gennari, and A. Wolisz, "Adaptation Algorithm for Adaptive Streaming over HTTP," 2012 19th Int. Pack. Video Work. (PV), Munich, pp. 173–178, 2012.

[7] T. C. Truong, D. Q. Ho, J. W. Kang, A. T. Pham, and S. Member, "Adaptive Streaming of Audiovisual Content using MPEG DASH," IEEE Trans. Consum. Electron., vol. 58, no. 1, pp. 78–85, 2012.

[8] H. T. Le, D. V. Nguyen, N. N. Pham, A. T. Pham, and T. C. Truong, "Buffer-based bitrate adaptation for adaptive HTTP streaming," Int. Conf. Adv. Technol. Commun., pp. 33–38, 2013.

[9] P. Juluri, V. Tamarapalli, and D. Medhi, "SARA: Segment aware rate adaptation algorithm for dynamic adaptive streaming over HTTP," 2015 IEEE Int. Conf. Commun. Work. ICCW 2015, pp. 1765–1770, 2015.

[10] T. H. Pham, A. D. Nguyen, T. Nguyen, H. T. Truong, and N. N. Pham, "Adaptation Method for Streaming of CBR Video over HTTP Based on Software Defined Networking," 2017 Int. Conf. Adv. Technol. Commun., no. 1, pp. 16–20, 2017.

[11] Y. Zhou, Y. Duan, J. Sun, and Z. Guo, "Towards simple and smooth rate adaption for VBR video in DASH," 2014 IEEE Vis. Commun. Image Process. Conf. VCIP 2014, pp. 9–12, 2015.

[12] T. Vu, H. T. Le, D. V. Nguyen, N. N. Pham, and T. C. Truong, "Future buffer based adaptation for VBR video streaming over HTTP," 2015 IEEE 17th Int. Work. Multimed. Signal Process. MMSP 2015, 2015.

[13] "Open Networking Foundation (ONF), "Software Defined Networking: the new norm for network"," White paper, 2012. .

[14] N. McKeown et al., "OpenFlow: Enabling Innovation in Campus Networks," ACM SIGCOMM Comput. Commun. Rev., vol. 38, no. 2, pp. 69–74, 2008.

[15] C. Cetinkaya, E. Karayer, M. Sayit, and C. Hellge, “SDN for Segment based Flow Routing of DASH,” 2014 IEEE Fourth Int. Conf. Consum. Electron. Berlin, pp. 74–77, 2014.
[16] D. L. C. Dutra, M. Bagaa, T. Taleb, and K. Samdanis, “Ensuring End-to-End QoS Based on Multi-Paths Routing Using SDN Technology,” 2017 IEEE Glob. Commun. Conf. GLOBECOM 2017 - Proc., vol. 2018-Janua, pp. 1–6, 2018.
[17] Y. S. Yu and C. H. Ke, “Genetic algorithm-based routing method for enhanced video delivery over software defined networks,” Int. J. Commun. Syst., vol. 31, no. 1, pp. 1–13, 2018.
[18] H. Nam, K. Kim, J. Y. Kim, and H. Schulzrinne, “Towards QoE-aware Video Streaming using SDN,” 2014 IEEE Glob. Commun. Conf., pp. 1317–1322, 2014.
[19] A. Bentaleb, A. C. Begen, and R. Zimmermann, “SDNDASH : Improving QoE of HTTP Adaptive Streaming Using Software Defined Networking,” ACM Multimed., pp. 1296–1305, 2016.
[20] A. Bentaleb, A. C. Begen, R. Zimmermann, and S. Harous, “SDNHAS: An SDN-enabled architecture to optimize QoE in HTTP adaptive streaming,” IEEE Trans. Multimed., vol. 19, no. 10, pp. 2136–2151, 2017.
[21] D. Bhat, A. Rizk, M. Zink, and R. Steinmetz, “Network Assisted Content Distribution for Adaptive Bitrate Video Streaming,” Proc. 8th ACM Multimed. Syst. Conf. - MMSys’17, no. June, pp. 62–75, 2017.
[22] E. Liotou, K. Samdanis, E. Pateromichelakis, N. Passas, and L. Merakos, “QoE-SDN APP: A Rate-guided QoE-aware SDN-APP for HTTP Adaptive Video Streaming,” IEEE J. Sel. Areas Commun., vol. 8716, no. c, pp. 1–17, 2018.
[23] S. . P. Le Callet, P.; Möller, “Qualinet White Paper on Definitions of Quality of Experience To cite this version : HAL Id : hal-00977812 Qualinet White Paper on Definitions of Quality of Experience Output from the fifth Qualinet meeting , Novi Sad ,” 2014.
[24] M. Seufert, S. Egger, M. Slanina, T. Zinner, T. Hossfeld, and P. Tran-gia, “A Survey on Quality of Experience of HTTP Adaptive Streaming,” Ieee Commun. Surv. Tutorials, vol. 17, no. 1, pp. 469–492, 2015.
[25] H. N. Nguyen, T. Vu, H. T. Le, N. N. Pham, and T. C. Truong, “Smooth quality adaptation method for VBR video streaming over HTTP,” 2015 Int. Conf. Comput. Manag. Telecommun. ComManTel 2015, pp. 184–188, 2016.
[26] S. Akhshabi, A. C. Begen, and C. Dovrolis, “An Experimental Evaluation of Rate-Adaptation Algorithms in Adaptive Streaming over HTTP,” Signal Process. Image Commun. 27(4), pp. 271–287, 2012.
[27] Wikipedia, “Round-robin scheduling,” 2018. .
[28] “Mininet.” [Online]. Available: http://mininet.github.io/. [Accessed: 25-Jun-2019].
[29] “Floodlight.” [Online]. Available: http://www.projectfloodlight.org/floodlight/. [Accessed: 25-Jun-2019].
[30] Orange Open Movie Project, “Elephants Dream,” 2006. .
[31] Cisco, “Comparing Traffic Policing and Traffic Shaping for Bandwidth Limiting.” [Online]. Available: https://www.cisco.com/c/en/us/support/docs/quality-of-service-qos/qos-policing/19645-policevsshape.html?dtid=osscdc000283. [Accessed: 10-Jul-2019].

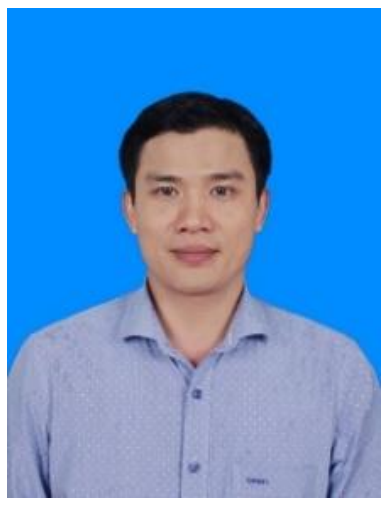

**Pham Hong Thinh** received the B.E. and M.E. degrees, both in Electronics and Telecommunication from the Hanoi University of Technology, Vietnam in 2002 and 2006, respectively. He is currently working toward a Ph.D. degree in Telecom. Engineering at Hanoi University of Science and Technology. From 2002 until now he has been working at Quy Nhon University, Vietnam. His research interests include Quality of Experience (QoE) optimization, Software Defined Networking (SDN), and content adaptation over HTTP.

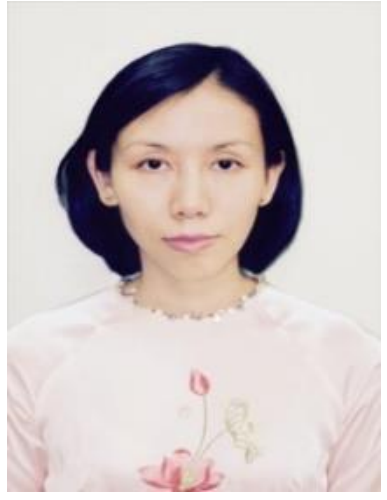

**Truong Thu Huong** received the B.Sc degree in Electronics and Telecommunications from Hanoi University of Science and Technology (HUST), Vietnam in 2001; the M.Sc. degree in Information and Communication Systems from Hamburg University of Technology, Germany in 2004; and Ph.D degree in Telecommunications from the University of Trento, Italy in 2007. She came back to work for Hanoi University of Science and Technology as lecturer in 2009 and became Associate Professor in 2018. Her research interest is oriented toward network security, artificial intelligence, traffic engineering in next generation networks, QoE/QoS guarantee for network services, green networking, development of Internet of Things ecosystems and applications.

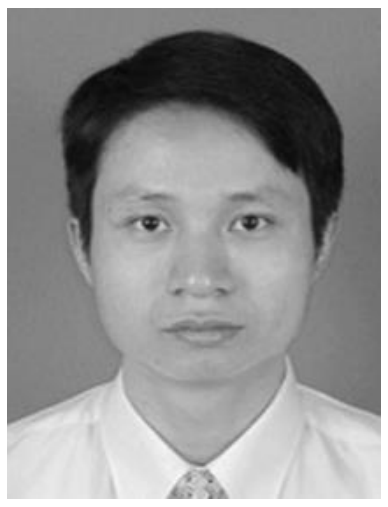

**Pham Ngoc Nam** received B. Eng. Degree In Electronics and Telecommunications from Hanoi University of Science and Technology (Vietnam) and M.Sc. degree in Artificial Intelligence from K.U. Leuven (Belgium) in 1997 and 1999, respectively. He was awarded a Ph.D. degree in Electrical Engineering from K.U.Leuven in 2004. From 2004 until 2018 he worked at Hanoi University of Science and Technology, Vietnam. And from August 2018, he has been working for the VIN University Project. His research interests include QoS management for multimedia applications, reconfigurable embedded systems and low-power embedded system design.

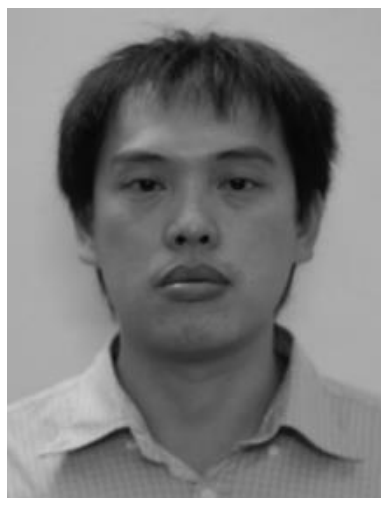

**Nguyen Huu Thanh** received BEng and MSc degrees in Electrical Engineering from Hanoi University of Science and Technology (Vietnam) in 1993 and 1995, respectively. He was awarded a PhD degree with summa cum laude in Computer Science from the University of Federal Armed Forces Munich (Germany) in 2002. In 2002 he joined Fraunhofer FOKUS in Berlin and worked on the areas of QoS-guarantees for multimedia in overlay networks. From 2004 until now he has been working for Hanoi University of Science and Technology, Vietnam. His research focuses on mobility, QoS/QoE and resource management of wireless broadband networks, new service platforms for future networks, Software-Defined Networking and the Future Internet.

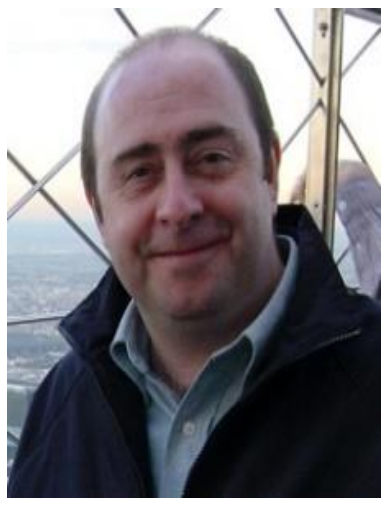

**Alan Marshall** holds the chair in Communications Networks at the University of Liverpool where he is director of the Advanced Networks Group. He is a senior member of IEEE and a Fellow of the IET. He has spent over 24 years working in the Telecommunications and Defence Industries. He has been visiting professor in network security at the University of Nice/CNRS, France, and Adjunct Professor for Research at Sunway University Malaysia. He has published over 200 scientific papers and holds a number of joint patents in the areas of communications and network security. He has formed a successful spin-out company Traffic Observation & Management (TOM) Ltd specialising in intrusion detection & prevention for wireless networks. His research interests include Network architectures and protocols; Mobile and Wireless networks; Network Security; high-speed packet switching, Quality of Service & Experience (QoS/QoE) architectures; and Distributed Haptics.